\documentclass[fleqn,12pt,twoside]{article}
\usepackage[headings]{espcrc1}

\readRCS
$Id: espcrc1.tex,v 1.2 2004/02/24 11:22:11 spepping Exp $
\usepackage{graphicx}
\usepackage[figuresright]{rotating}

\newcommand{\AmS}{{\protect\the\textfont2
  A\kern-.1667em\lower.5ex\hbox{M}\kern-.125emS}}

\title{Collection of CERES Results}

\author{D.~Mi{\'s}kowiec
(for the CERES Collaboration)\footnote{
For the full CERES Collaboration author list and acknowledgments, 
see appendix 'Collaborations' of this volume.}\\
Gesellschaft f{\"u}r Schwerionenforschung mbH, 
Planckstr. 1, D-64291 Darmstadt}
       
\runauthor{D.~Mi{\'s}kowiec}

\def\minv{$m_{ee}$}

\def\pt{$p_t$}
\def\ee{e$^+$e$^-$}
\def\runinfo{The data are from central Pb+Au at 158~GeV per nucleon.}
\begin{document}

\maketitle

\begin{abstract}
The measurement of central Pb+Au collisions at the maximum SPS energy 
of 158 GeV per nucleon in the fall of 2000 was the first run of the fully 
upgraded CERES and at the same time the last run of this experiment. 
Today, after several years of tedious calibration, the physics analysis 
is in its peak. 
A snapshot of the current results is presented here. 
\end{abstract}

\section{Introduction}
CERES is a dilepton experiment at the CERN SPS, known for its observation 
of enhanced production of low mass \ee\ pairs in collisions between heavy 
nuclei \cite{ceres-sau-pbau}. 
The fact that the excess pairs had invariant masses above approximately 
twice the pion mass 
and that the enhancement was absent in proton induced collisions 
\cite{ceres-p} pointed towards pion annihilation $\pi\pi\to \rho \to e^+e^-$ 
as the additional source of lepton pairs. 
To fully account for the measured spectrum, however, the shape of the $\rho$ peak 
needs to be modified. 
Theoretical models expect a shift of the peak  
to lower masses \cite{brown} or a broadening and a slight shift up 
\cite{rapp} when $\rho$ is immersed in high density hadronic matter. 
Testing the predictions of (and possibly distinguishing between)  
the two scenarios was among the objectives of the CERES upgrade. 
In this note I list several results of the ongoing analysis 
of the data from the last run of the upgraded CERES. 
For the sake of limited space two of the topics presented in the talk, 
the elliptic flow of $\Lambda$ and the back-to-back correlations, 
are omitted. 
These results are described in detail in \cite{jovan} and \cite{mateusz}, 
respectively. 

\section{Experiment}
\begin{figure}[t]
\vspace{-5mm}
\rotatebox{270}{\scalebox{0.57}{\includegraphics{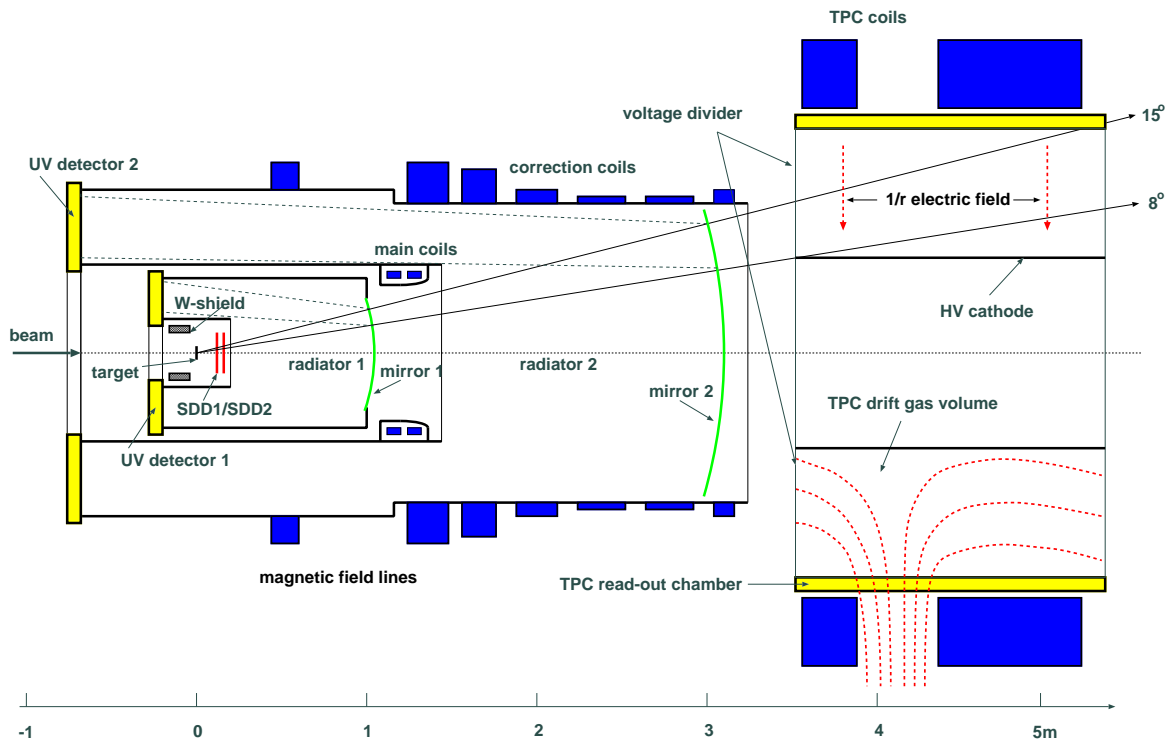}}}
\caption{
The upgraded CERES setup in 2000. 
The apparatus has a cylindrical symmetry. 
The beam enters from the left. 
The silicon detectors (SDD) give tracking and vertex reconstruction, 
and the Ring Imaging Cherenkov detectors (RICH) electron identification. 
The Time Projection Chamber (TPC) measures the momentum and the energy loss.}
\label{fig:setup}
\end{figure}

The upgrade of CERES in 1997-1998 by a radial Time Projection Chamber (TPC) 
allowed to improve the momentum resolution and the particle identification 
capability while retaining the cylindrical symmetry. 
The TPC also opened the possibility of measuring hadrons. 
The upgraded experiment is shown in Fig.~\ref{fig:setup}. 
Charged particles emitted from a segmented Au target 
first pass through two silicon drift detectors (SDD) located 100~mm and 
138~mm from the target.  
The two detectors are used to reconstruct angles of the charged particle 
tracks with $\Delta \theta$=0.2~mrad and $\Delta \phi$=2~mrad, and the interaction vertex 
with $\Delta z = 200$~$\mu$m, the coordinate $z$ being along the beam axis. 
Subsequently, the particles traverse two RICH detectors with 
$\gamma_{\rm THR}$=32 which serve for electron identification. 
In the upgraded CERES the magnetic field between the two RICHes 
is switched off and thus the rings in the two RICHes are aligned 
which leads to a better particle identification.  
Finally, the particles pass through the newly built TPC. 
The ionization electrons drift outward towards the 16 readout chambers. 
The electric field is inversely proportional to the radius and thus 
the drift is fastest where the hit density is highest. 
The semi-radial magnetic field deflects the particles in the azimuthal 
direction. 
The momentum resolution reached after the final calibration is 
\begin{equation}
\frac{\Delta p}{p} = 2\% \oplus 1\% \cdot p/{\rm GeV} 
\end{equation}
resulting in $\frac{\Delta m}{m} = 0.038$ for the $\phi$ meson in the \ee\ channel. 
Independent particle identification is achieved using the energy loss 
in the TPC gas with $\frac{\Delta (dE/dx)}{(dE/dx)} = 0.10$. 

The data presented here come from the run in the fall 2000. 
In this run about 30 million Pb+Au collision events at 158 GeV per nucleon 
were collected, most of them with centrality of top 7\% of geometrical cross 
section. 
Small samples of the 20\% and the minimum bias collisions, as well as a short 
run at 80 AGeV, were recorded in addition. 

\section{Centrality and charged particle multiplicity}
\begin{figure}[t]
\hspace*{3cm}\scalebox{0.6}{\includegraphics{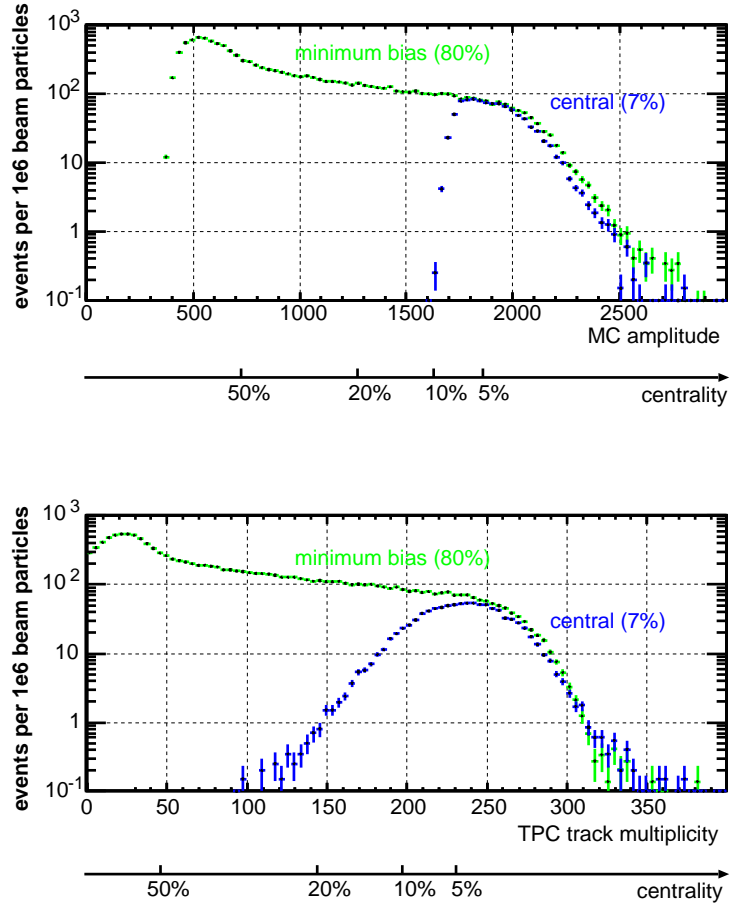}}
\vspace{-5mm}
\caption{Distributions of the 
pulse height of the MC scintillator detector (top) 
and the TPC track multiplicity (bottom) 
used for centrality determination. 
The MC detector was used in the trigger. 
The distributions shown are before the run-by-run correction. \runinfo}
\label{fig:multiplicity}
\end{figure}

The collision centrality was determined via the charged particle multiplicity 
around midrapidity $y_{\rm beam}$/2=2.91. 
Two variables, the amplitude of the Multiplicity Counter (MC) 
(single scintillator covering 2.3$<\eta<$3.5) and the track multiplicity in the 
TPC (2.1$<\eta<$2.8), were alternatively used as the centrality measure 
(Fig.~\ref{fig:multiplicity}). 
Knowing the data acquisition dead time factor and the target thickness, 
and assuming that all beam particles were hitting the target, the event 
counts can be translated to the cross section for collisions with a given 
multiplicity. 
The integrated cross section, divided by the geometrical cross section 
$\sigma_{G}=6.94$ barn, is shown as the additional axis in 
Fig.~\ref{fig:multiplicity}. 

\begin{figure}[t]
\hspace*{2cm}\scalebox{0.6}{\includegraphics{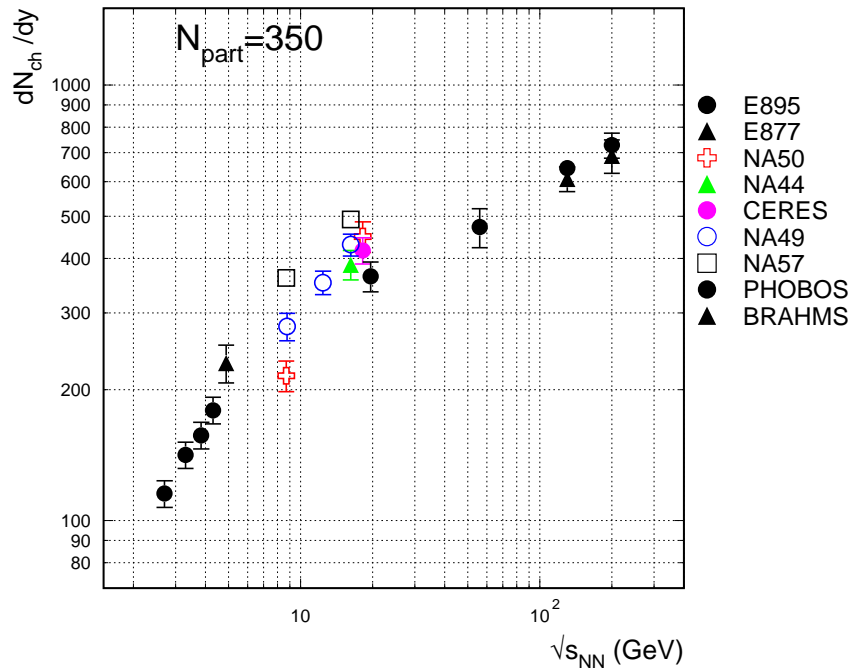}}
\caption{Beam energy dependence of the rapidity density of charged 
particles \cite{anton}. 
The CERES point at 158 GeV per nucleon ($\sqrt{s}_{NN}$=17.2~GeV)
is in agreement with the results of the other SPS experiments. }
\label{fig:anton}
\end{figure}

Charged particle multiplicity within 2$<\!\eta\!<$3 was determined from 
coincidences between the hits in the two silicon detectors. 
Random coincidences, estimated by rotating one of the detectors, 
were subtracted. 
The finite two-hit resolution problem was solved by requiring 
certain minimum separation between every pair of hits $\Delta r$ and 
extrapolating to $\Delta r=0$. 
The $\delta$-electron contribution was estimated from a run with beam trigger 
and subtracted. The resulting pseudorapidity density of charged 
particles per participant, 
extrapolated to $\eta=3.1$ where the maximum is expected to be, 
is d$N_{ch}$/d$\eta$(3.1)/$N_{\rm part}$=1.18$\pm$0.15 (syst.),  
independent of centrality within 0-50\%. 
The number of participants for a given centrality is calculated 
using the nuclear overlap model \cite{overlap}. 
The central collision d$N_{ch}$/d$y$(2.9), 
estimated via 1.02 times d$N_{ch}$/d$\eta$(3.1), 
fits the available beam energy systematics (Fig.~\ref{fig:anton}). 

\section{Dileptons}
\begin{figure}[h]
\begin{minipage}[t]{30mm}
\hspace*{-6mm}
\scalebox{0.44}{\includegraphics{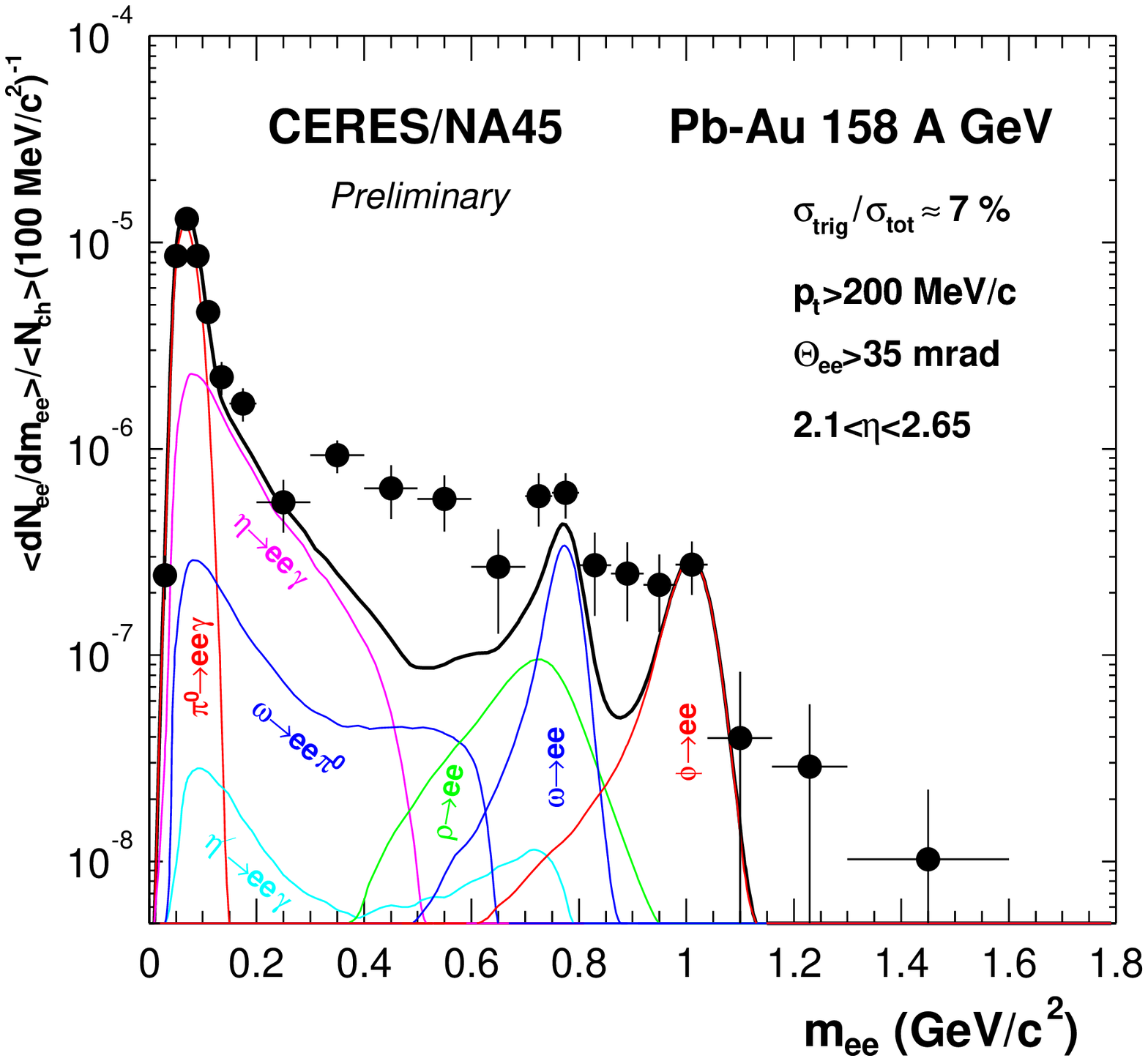}}
\end{minipage}
\hspace{\fill}
\begin{minipage}[t]{75mm}
\scalebox{0.44}{\includegraphics{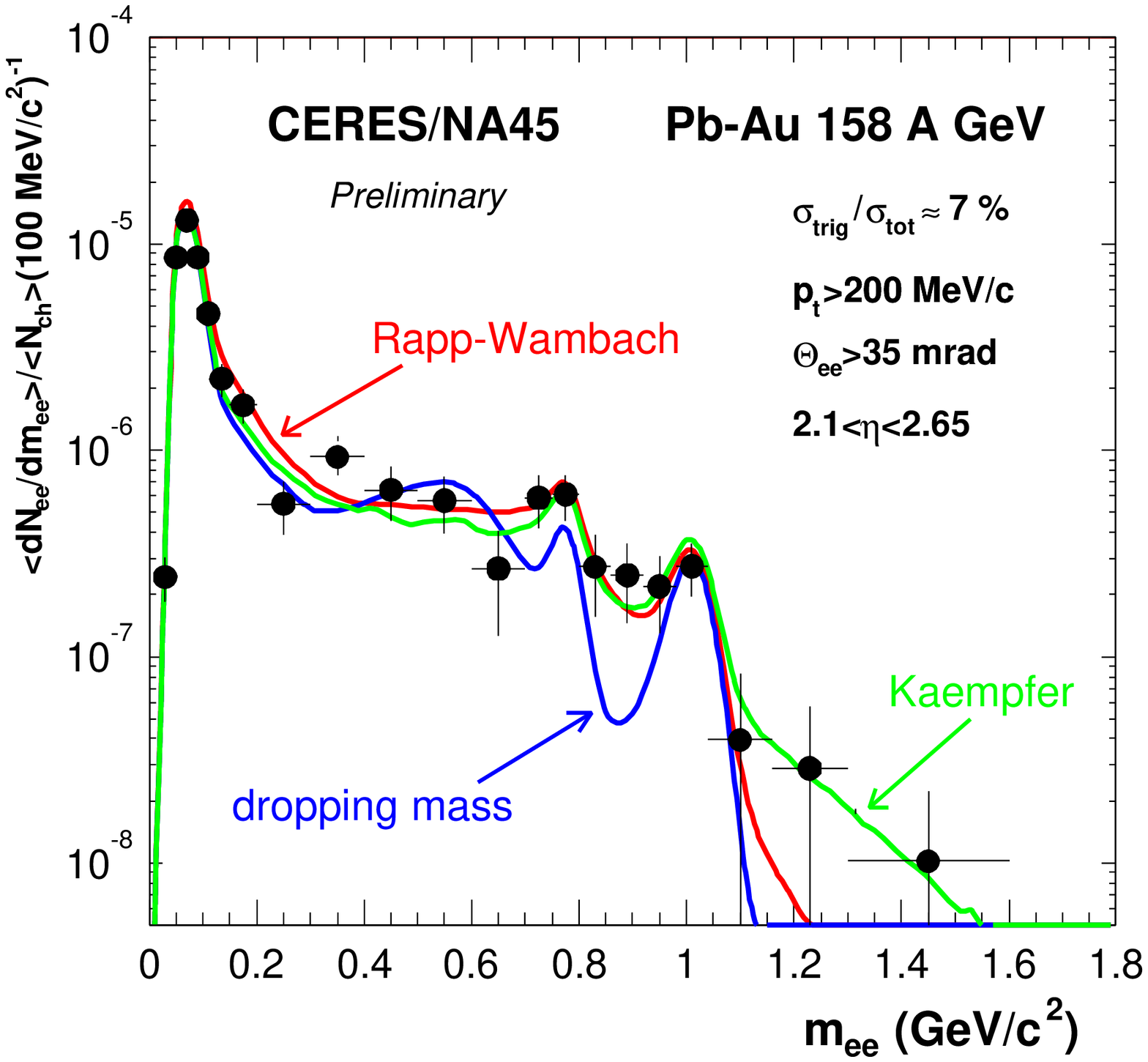}}
\label{fig:mass_pt200}
\end{minipage}
\vspace{-1cm}
\caption{Dielectron mass spectrum with the hadron decay cocktail (left) 
and with the models (right). 
The spectrum contains 2571$\pm$224 \ee\ pairs with \minv$>$0.2~GeV/c$^2$, 
with the signal to background ratio of 1:21. 
The multiplicity of charged particles, averaged over event centralities 
and over the acceptance, is $\langle$d$N_{ch}/$d$\eta\rangle$=335. See the text for 
description of the models in the right panel. }
\end{figure}

Electron tracks are identified by requiring a ring in the combined RICH 
detectors and a high $dE/dx$ in the TPC. 
The combined electron efficiency and the pion suppression factor are 68\% and 
4$\cdot 10^4$, respectively, for $p$=1.5~GeV/c. 
The \ee\ pairs from $\pi^0$ Dalitz decay and from $\gamma$ conversion are 
recognized via their small opening angle and marked such that they are 
not paired with other electrons. 
The combinatorial background is removed from the \ee\ mass distribution 
by subtracting the like-sign or the event-mixing spectrum. 
The resulting signal spectrum is corrected for efficiency, determined 
in a Monte Carlo simulation and parametrized in terms of a 
single track efficiency. The latter depends primarily on the 
hit density and varies between 0.2 and 0.5 for 0.14$<$$\theta$$<$0.24. 
The efficiency corrected and absolutely normalized mass spectrum 
exhibits an enhancement of \ee\ pairs in the mass range between 
0.2 and 0.6~GeV/c$^2$ by a factor of 2.8$\pm$0.5 (stat). 
This is shown in the left panel of Fig.~\ref{fig:mass_pt200}. 
The systematic error of the overall normalization is 21\%. 

The three model calculations, compared to the data in the right panel 
of Fig.~\ref{fig:mass_pt200}, are based on three different scenarios. 
The line labelled with ``Rapp-Wambach'' represents a calculation 
including a modification of the $\rho$ spectral function by hadronic 
medium \cite{rapp}. 
The ``dropping mass'' line is a result of a calculation by R.~Rapp 
in which the $\rho$ meson is shifting according to the Brown-Rho 
scaling \cite{brown}. 
Finally, a thermal emission calculation gives the line labelled as 
``Kaempfer'' \cite{kaempfer}. 
The calculations were done for the actual centrality and 
agree reasonably well with the experimental data 
except for the ``dropping mass'' curve which in the region between 
$\rho$ and $\phi$ exhibits a minimum not present in the data. 

\section{Resolution of the $\phi$ puzzle}
\begin{figure}[h]
\vspace{-1mm}
\begin{minipage}[t]{0mm}
\hspace*{-5mm}\scalebox{0.45}{\includegraphics{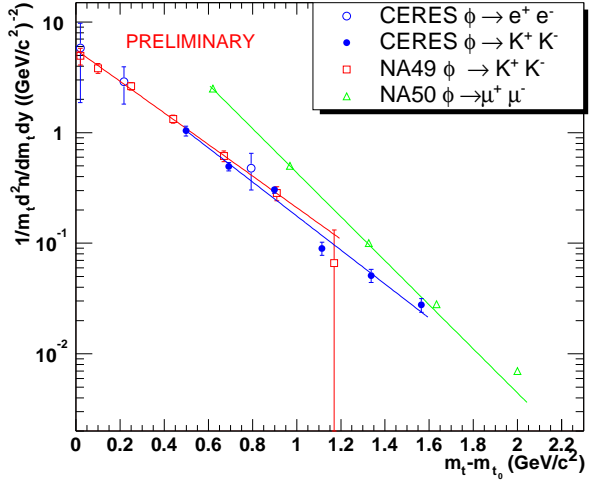}}
\end{minipage}
\hspace{\fill}
\begin{minipage}[t]{65mm}
\vspace{-5.5cm}
\caption{Transverse mass spectra of $\phi$ reconstructed in the 
K$^+$K$^-$ and the \ee\ decay channels by CERES, compared to the results 
of NA49 and NA50. No indication of discrepancy between the hadronic 
and leptonic decay channels is visible in the CERES data. }
\label{fig:ana-phipt}
\end{minipage}
\end{figure}

The discrepancy between the $\phi$ measured in the KK and $\mu\mu$ channels 
respectively by NA49 and NA50, pointed out in \cite{roehrich}, 
might indicate a possible medium effect on the $\phi$ decay or on its 
reconstruction probability in the experiment. 
Indeed, short-lived mesons decaying to hadrons inside the fireball may 
get reconstructed with a wrong invariant mass, or not reconstructed at all, 
because of the rescattering of the decay products. 
The vacuum lifetime of the $\phi$ meson, however, is too long for this 
effect to explain the observed difference between the hadronic and leptonic 
channels \cite{johnson}. 
The upgraded CERES is capable of measuring hadrons and thus reconstructing 
$\phi\to K^+K^-$. 
The $\phi\to$\ee\ channel was evaluated by integrating the absolutely normalized 
dielectron mass spectrum between 0.9 and 1.1 GeV/c$^2$, and reducing the 
obtained number by 28\% to account for the possible high mass tail of the 
$\rho$ meson. 
The $p_t$-spectra of $\phi$ reconstructed in these two ways agree 
within the error bars and are consistent with the result obtained by NA49 
(Fig.~\ref{fig:ana-phipt}). 

\section{Pion-proton correlations}
\begin{figure}[h]
\begin{minipage}[t]{100mm}
\vspace{-3mm}
\hspace{-5mm}
\scalebox{0.4}{\rotatebox{270}{\includegraphics{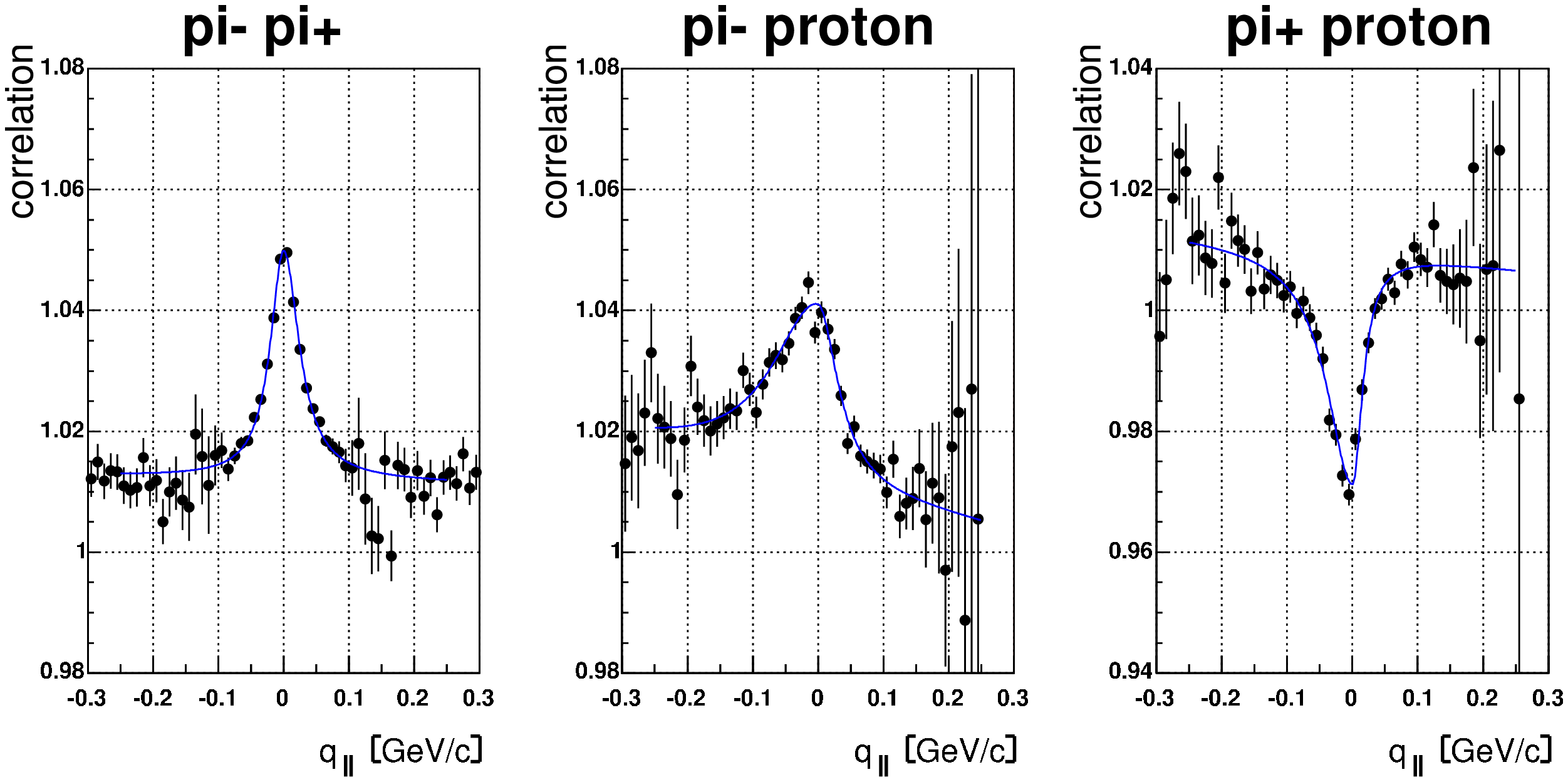}}}
\vspace{-9mm}
\caption{Non-identical particle correlations as functions of the 
relative momentum along the pair velocity $q_{\|}$ (in the pair c.m. frame). 
The pair \pt\ is 0.4-1~GeV/c for $\pi\pi$ and 1-2~GeV/c for $\pi$p. 
\runinfo}
\label{fig:darek-cor-exa}
\end{minipage}
\hspace{\fill}
\begin{minipage}[t]{50mm}
\vspace{0mm}
\hspace{-6mm}
\scalebox{0.27}[0.27]{\includegraphics{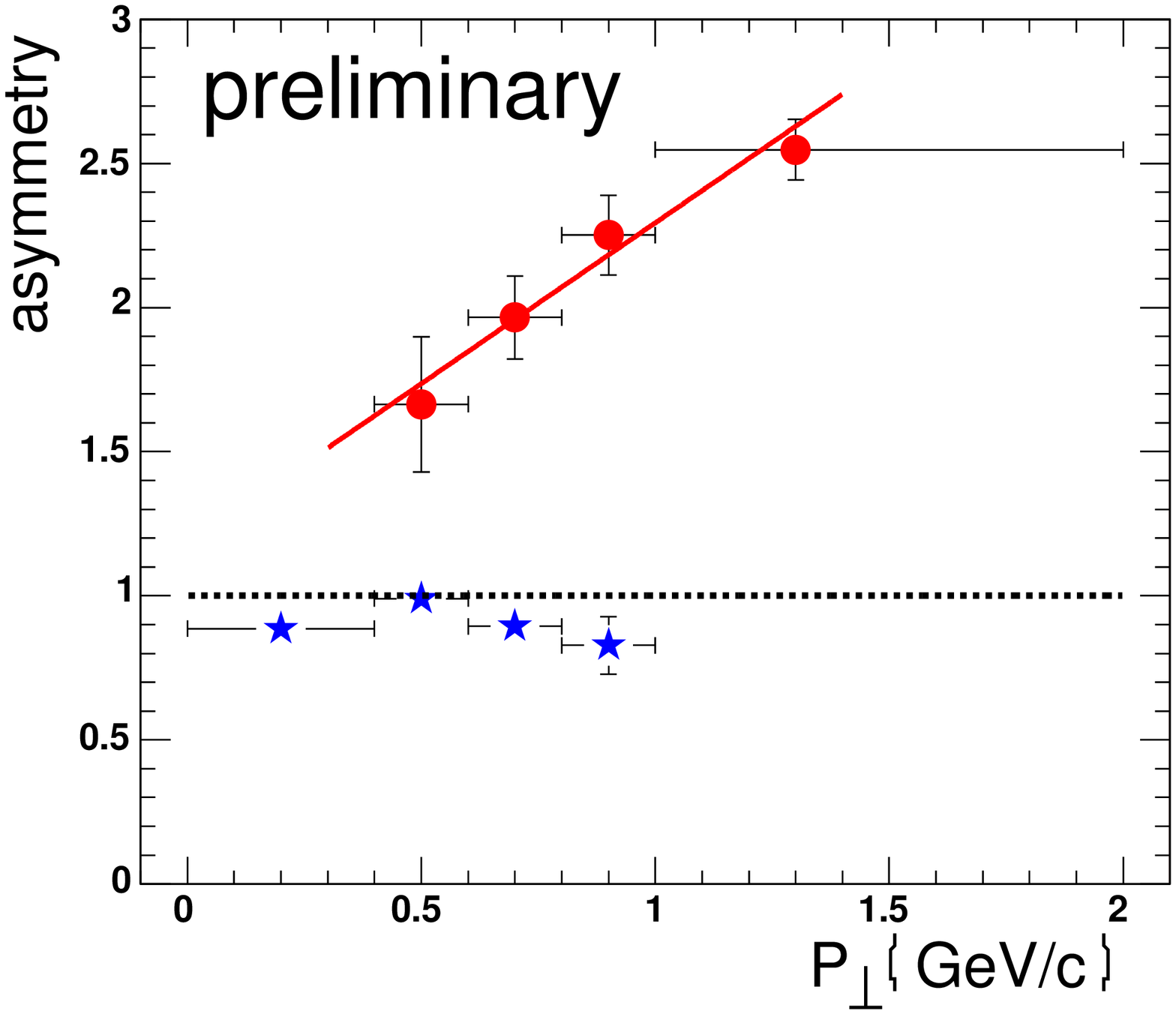}}
\vspace{-9mm}
\caption{Asymmetry of the $\pi$p (dots) and 
the $\pi^+\pi^-$ (stars) correlations vs. pair transverse momentum.}
\label{fig:darek-cor-pt}
\end{minipage}
\end{figure}

The shape of non-identical particle correlation functions $C({\bf q})$, 
${\bf q}:={\bf p_2}-{\bf p_1}$,  reflects the shape of 
the relative source distributions $S(r_2^\mu-r_1^\mu)$. 
Particularly, a difference between the average freeze-out position or time 
of two particle species reveals itself as an asymmetry of the 
correlation function at small $q$ \cite{lednicky}. 
Slices of $\pi^-\pi^+$, $\pi^-$p, and $\pi^+$p correlation functions 
are shown in Fig.~\ref{fig:darek-cor-exa}. 
The structures at low $q$ come from the mutual Coulomb interaction. 
The asymmetry in the pion-proton correlations 
indicates that the proton source is located at a larger radius 
than the pion source, or that protons are emitted earlier than 
pions. 
The asymmetry can be conveniently parametrized by fitting a 
Lorentz curve, which happens to match the shape, 
separately to the left and to the right half of the peak, 
and taking the ratio of the two widths. 
A pair generator with a 6.5~fm displacement between the sources 
of protons and pions can reproduce the asymmetry observed for pairs with 
a \pt\ within 0.6-2~GeV/c. 
A detailed \pt\ dependent analysis is in progress 
(Fig.~\ref{fig:darek-cor-pt}). 

\section{Transverse momentum fluctuations}
\begin{figure}[h]
\vspace{-11.5cm}
\hspace*{8.9cm}
\begin{minipage}[t]{6.9cm}
\vspace{-0.2mm}
\scalebox{0.6}[0.46]{\includegraphics*{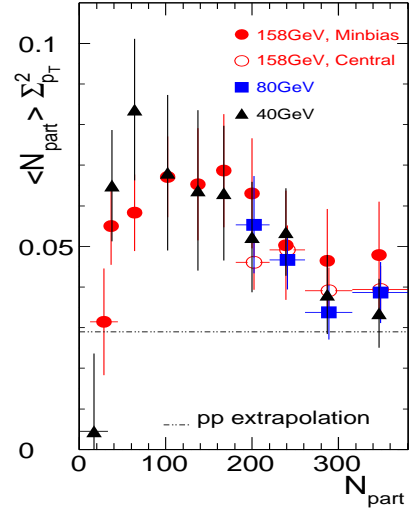}}
\caption{Fluctuations of mean \pt\, scaled with the number of participants, 
versus centrality. }
\label{fig:georgios0}
\end{minipage}
\end{figure}

\begin{figure}[h]
\vspace*{-9mm}
\begin{minipage}[t]{75mm}
\hspace{-5mm}\scalebox{0.4}{\rotatebox{0}{\includegraphics{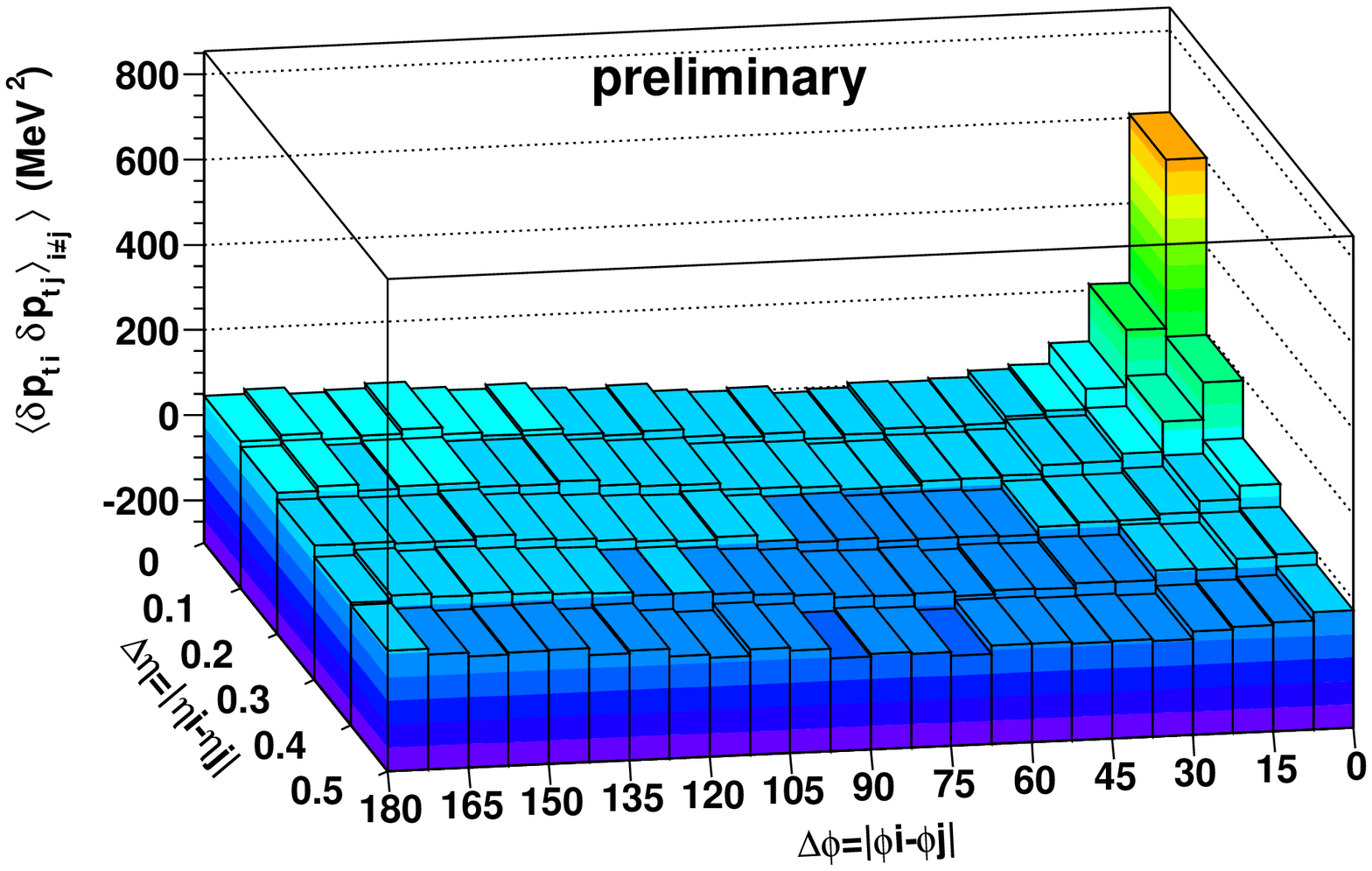}}}
\vspace{-1cm}
\caption{Transverse momentum covariance between pairs of tracks for 
different pair opening angles. \runinfo}
\label{fig:georgios1}
\end{minipage}
\hspace{\fill}
\begin{minipage}[t]{75mm}
\vspace{-5.45cm}
\hspace{-7mm}
\scalebox{0.75}[0.75]{\rotatebox{0}{\includegraphics{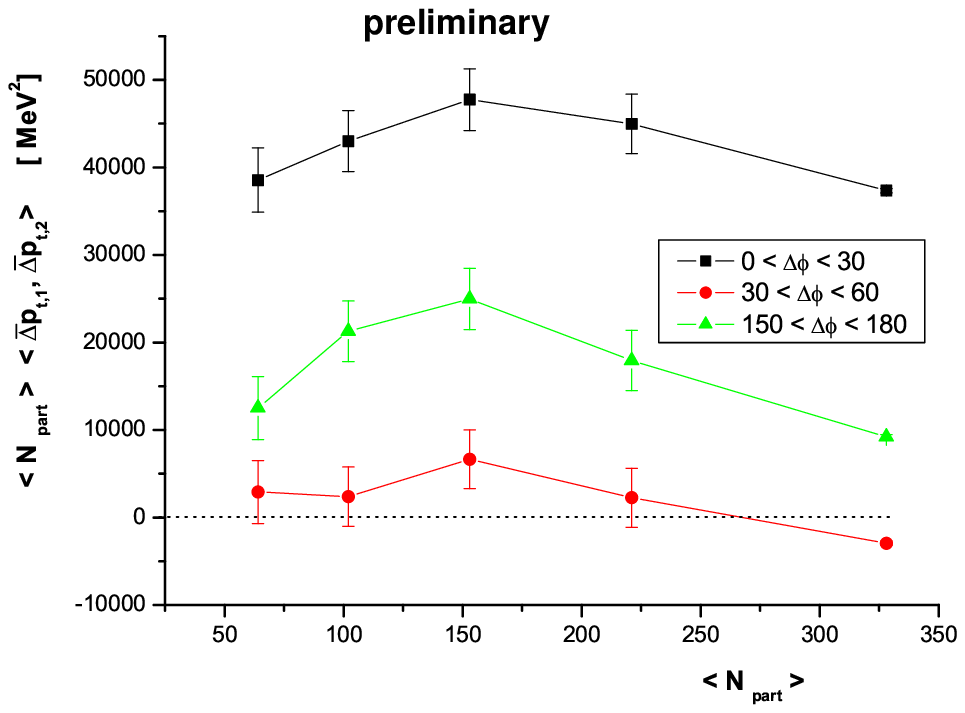}}}
\vspace{-1.5cm}
\caption{Centrality dependence of the transverse momentum covariance 
for three regions of $\Delta\phi$. }
\label{fig:georgios2}
\end{minipage}
\end{figure}

\noindent\begin{minipage}[t]{8.cm}
Enhanced event-by-event fluctuations of transverse momenta are 
considered to be one of the signatures of the vicinity to the 
critical point of the QCD phase transition. 
Unfortunately, the fluctuations observed so far turned out 
to be rather independent of beam energy \cite{ceres-pt}. 
On the other hand, the fluctuations seem to be a non-monotonic 
function of centrality (Fig.~\ref{fig:georgios0}). 
The variable $\Sigma_{pt}^2$ used there is the difference between 
the variances of the $\langle$\pt$\rangle$ (events) and \pt\ (tracks) 
distributions, 
scaled such to be zero in the case of independent particle emission 
\cite{voloshin1}, 
and normalized to the mean \pt. 

More insight into the origin of the observed fluctuations is gained 
by studying the transverse momentum covariance 
$\langle\langle\delta p_{t_i}\delta p_{t_j}\rangle_{i\not= j}\rangle$ \cite{voloshin2} between pairs 
of tracks of a given 
opening angle. 
The covariance map (Fig.~\ref{fig:georgios1})  

\end{minipage}

\hspace*{-4mm}reveals 
several structures and demonstrates that if the fluctuations were to 
be characterized by one single number the 
result would depend on the experiment's acceptance. 
The landscape is dominated by the short range correlation peak at small 
opening angles, most probably originating from Bose-Einstein and 
Coulomb effects between pairs of particles emitted with similar velocities,  
and the broad maximum at $\Delta\phi$=180$^{\rm o}$ which contains back-to-back  
correlations like those observed at RHIC and the SPS \cite{ceres-jets}. 
The elliptic flow introduces a cos(2$\Delta\phi$) modulation. 
The slight decrease with increasing $\Delta\eta$, 
which can be reproduced in event mixing, 
comes probably from the \pt($\eta$) dependence. 

The overall fluctuations thus seem to be dominated by the short range 
and the away-side two-particle correlations which are not expected 
to be sensitive to the critical point. 
The non-monotonic centrality dependence of the overall fluctuations is 
indeed visible in the separate analysis of these two components 
(two upper sets of points in Fig.~\ref{fig:georgios2}). 
Since the critical point fluctuations should be present for all 
opening angles the best strategy seems to be to focus on the 
fluctuations in the region of $30^{\rm o}<\Delta\phi<60^{\rm o}$, free 
of the influence of the two mentioned components, and where the 
elliptic flow does not matter. 
The \pt\ fluctuations in this region turn out to be close to zero 
and independent on centrality within the error bars 
(lowest set of points in Fig.~\ref{fig:georgios2}). 
It would be interesting to analyze the beam energy dependence of this 
quantity. 

\section{Summary}
The data of the upgraded CERES run confirms the existence of 
a dilepton excess in the \minv\ range between the $\pi^0$ and 
the $\rho$. 
With the improved mass resolution the data disfavors  
the dropping $\rho$-meson mass scenario in its present implementation. 

The discrepancy between the NA49 and NA50 measurements of the 
hadronic and leptonic decays of the $\phi$ meson is not manifest 
in the CERES data. 
The \ee\ and K$^+$K$^-$ decay channels are mutually consistent and are 
in agreement with the NA49 measurement. 

The distributions of pions and protons at freeze-out are either 
displaced by about 6~fm or these two particle species are emitted, 
on average, at different times. 
This is the result of an analysis of the Coulomb correlations 
between pions and protons. 

A differential analysis identifies the short range correlations 
and the away-side correlations as the dominant source of the 
event-by-event \pt\ fluctuations. 
The pairs emitted with $\Delta \phi$=45$^{\rm o}$, 
which might provide a direct probe of the critical point, 
do not contribute significantly to the fluctuations. 


\end{document}